\documentclass[amsmath,amsfonts,twocolumn,superscriptaddress,prb]{revtex4}
\usepackage{graphicx}
\usepackage{epsfig}
\usepackage{bbm}
\usepackage{bm}
\usepackage{dcolumn}
\usepackage{amsmath}
\usepackage{amssymb}
\usepackage{color}
\usepackage[varg]{txfonts}

\newcommand{\beg}{\begin{equation}}
\newcommand{\en}{\end{equation}}
\newcommand{\bp}{\mathbf p}
\newcommand{\bq}{\mathbf q}
\newcommand{\bk}{\mathbf k}
\newcommand{\br}{\mathbf r}

\newcommand \bel  {\begin{align}}
\newcommand \enl  {\end{align}}

\newcommand{\veps}{\varepsilon}
\newcommand{\eps}{\epsilon}

\newcommand{\up}{\uparrow}
\newcommand{\dn}{\downarrow}
\newcommand{\dg}{^\dagger}

\newcommand{\St}{\mathrm{St}\,}
\def\XXint#1#2#3{{\setbox0=\hbox{$#1{#2#3}{\int}$}
     \vcenter{\hbox{$#2#3$}}\kern-.5\wd0}}

\begin{document}

\title{Magnetic Field Induced Nonlinear Transport in LaTiO$_3$/SrTiO$_3$ Interfaces}

\author{Aidan Steineman}
\affiliation{Department of Physics, Kent State University, Kent, Ohio 44242, USA}

\author{Maxim Khodas}
\affiliation{The Racah Institute of Physics, The Hebrew University of Jerusalem, Jerusalem 91904, Israel}

\author{Maxim Dzero}
\affiliation{Department of Physics, Kent State University, Kent, Ohio 44242, USA}

\date{\today}

\begin{abstract}
Motivated by the recent experimental measurements of the nonlinear longitudinal resistance of the spin-orbit coupled electron gas in the (111) LaTiO$_3$/SrTiO$_3$ interfaces under external in-plane magnetic field [G. Tuvia \emph{et al.}, Phys. Rev. Lett. 132, 146301 (2024)], we
formulate a theory of nonlinear electronic transport based on the analysis of the quantum kinetic equation for the Wigner distribution function. Specifically, we evaluate the magnetic field dependence of the second harmonic of the current density at arbitrary values of the magnetic field. The magnitude of the second harmonic increases linearly with the magnetic field at small fields. Upon further increase of the magnetic field, the second harmonic response reaches its maximum value. We find that the position of the peak and its width strongly depend on the relaxation rate due to disorder.  Importantly, we discover that the direction of the nonlinear contribution to the current can be completely reversed when the magnetic field reaches a certain critical value.
\end{abstract}

\maketitle

\section{Introduction}
The physical phenomena observed in perovskite-based heterostructures have attracted quite significant attention from both experimental and theoretical condensed matter communities during the past two decades.\cite{OxideInterfaces-Review,OxideInterfaces-2023} In particular, two-dimensional interfaces formed in La$X$O$_3$/SrTiO$_3$  heterostructures (with $X$=Al,Ti) provide a fruitful playground for exploring new physical effects such as unusual magnetoresistance and spin texture driven by the electron-electron correlations and spin-orbit coupling.\cite{SrTi03,SOCoupling,Rashba-SO,BMER-2018,Lesne-2023,MR-2017,Non-reciprocal2019} In addition, in the case of  LaAlO$_3$/SrTiO$_3$, a spin-orbit coupled electronic liquid formed at the interface may host an unconventional superconductivity while, in the interface formed between a Mott insulating material LaTiO$_3$ and a band insulator SrTiO$_3$, intrinsic electronic correlations may host the electronic states with unconventional orbital and magnetic order.\cite{Dagan-2010SC,Dagan-2010,Dagan-Link2017,Altman2012,Altman2013} Electrons confined at the LaTiO$_3$/SrTiO$_3$ interface also develop superconductivity at temperatures $T_c\approx300$ mK.\cite{Biscaras2010}

Recent advances in exploring the nonlinear transport in polar metals and superconductors \cite{SH2017,Non-reciprocal2019,Rectify1,FDoubling2021,NR-Transport,NR-Transport2,PlanarHall2021,NonlinearMT2019,CIMOKE,Spivak2009,Sodemann2015,BerryDipole1,BerryDipole2,BerryDipole3,BerryDipole4,BerryDipole5} has also prompted the investigation of nonlinear transport properties at the metallic interfaces in oxide heterostructures. \cite{Nonlinear-Interface2019}
Most recently, Tuvia \emph{et al.} \cite{Dagan:2024} have studied the electronic properties of the (111) LaTiO$_3$/SrTiO$_3$ interface by measuring the longitudinal part of the nonlinear resistivity tensor $R_{xx}\propto \sigma_2/\sigma_{\textrm{D}}^2$, where $\sigma_{\textrm{D}}$ is the Drude conductivity and $\sigma_2$  is the second harmonic contribution to the conductivity which can be obtained from the components of the tensor $\chi_{\textrm{abc}}$ defined by
\beg\label{SecondHarmDefine}
j_{\textrm{a}}(\omega)=\chi_{\textrm{abc}}(\omega)E_{\textrm{b}}(\omega)E_{\textrm{c}}(\omega).
\en
In order to generate a local nonlinear response, the constant magnetic field has been applied parallel to the plane of the interface. It was observed that the second harmonic response increases with an increase in the magnitude of the external field. As the value of the magnetic field reaches some critical value, the second harmonic exhibits a peak. This observation has been explained using the simple model of two Rashba spin-orbit split bands by noting that  the maximum in the second harmonic appears when the magnetic field is such that the bands cross exactly at the chemical potential. If an out-of-plane magnetic field is applied, the effect becomes drastically suppressed due to the gap opening at the Dirac point. 

In this paper, we will use the same model as the one used in Ref. \onlinecite{Dagan:2024} to demonstrate that  the situation is more nuanced. Specifically, we will show that the second harmonic response exhibits a strong dependence on disorder which determines both the position and width of the peak. Importantly, we will demonstrate that with an increase in the magnetic field the second harmonic first goes to zero and then changes its sign.

As we have mentioned above, we consider an electron gas in two dimensions and in the presence of a weak disorder potential. In the absence of the inversion symmetry, accounted for by including the Rashba spin-orbit interaction, it is clear that the second harmonic contribution to the current density necessarily requires another vector to determine its direction. Indeed, the second harmonic of the current is a vector quantity and, in the problem that we consider, is proportional to the square of electric field. At first 
glance, the two-dimensional electronic system that we discuss provides such a vector: normal to the plane of motion. However, this scenario has to be discarded since the motion of electrons is confined to a plane. Therefore, one necessarily requires either the gradient  or static in-plane magnetic field to play a role of such a vector. We should also mention that in the case of non-zero Berry curvature and in the absence of the mirror symmetry, the Berry curvature dipole lies in the plane of motion and therefore may also serve as a purely local source of second harmonic generation.\cite{Sodemann:2015} For the model which we will consider below, the mirror symmetry is preserved and so the only source of the local second harmonic response is a combination of an external in-plane magnetic field ${\mathbf B}$ and nonzero spin-orbit coupling.  

Generally, for the case that we consider below there are three possible vector combinations which determine the direction of the second harmonic response: $({\vec e}_z\times{\mathbf B}){E}^2$, $[({\mathbf B}\times{\mathbf {E}})\cdot{\vec e}_z]{\mathbf {E}}$ and 
$({\mathbf E}\cdot{\mathbf B})({\vec e}_z\times{\mathbf E})$ (here ${\vec e}_z$ is a unit vector orthogonal to the plane of motion and ${\mathbf E}$ is an external electric field).\cite{Dzero:2024}  
In passing, we note that at arbitrary order in $B$, one has more combinations: 
$\bm{j} = ({\vec e}_z \times \bm{B})[a_B (\bm{E}\cdot \bm{B})^2 + b_B (\bm{E}\times \bm{B})^2]$, and also $\bm{j} = c_B \bm{B} (\bm{B}\cdot\bm{E})[{\vec e}_z\cdot (\bm{E} \times \bm{B})]$.
Here $a_B$, $b_B$, and $c_B$ are arbitrary functions of $B^2$. One can prove that there is nothing else {\it to all} orders in $B$.

It is straightforward to show that out of these three vectors only two are linearly independent. In the limit when the magnetic field is weak, one immediately concludes that the magnitude of the second harmonic will increase linearly with the magnetic field. It is important to emphasize that the $ac$-second harmonic response is resonant in nature at frequencies comparable with the energy splitting between the two spin-orbit split bands.\cite{Dzero:2024} The width and the amplitude of the corresponding resonance is determined by the dimensionless parameter $\omega\tau$, where $\omega$ is the frequency of the electric field and $\tau^{-1}$ is the disorder scattering rate.  

On the other hand, one may also expect that at some critical value of the external magnetic field $B_{\textrm{c}}$, the second harmonic response will decrease with the increase in the values of the magnetic field. This is due to the fact that the application of a magnetic field leads to a relative shift in the position of the spin-orbit split bands. At some value of the field, the bands which at zero field cross at the $\Gamma$-point of the two dimensional Brillouin zone, will cross exactly at the chemical potential giving rise to a two-dimensional Dirac cone. By virtue of the kinematic constraints imposed by the arguments of the single-particle distribution function the second harmonic must become vanishingly small at some value of the magnetic field. We thus arrive to the conclusion that the second harmonic must exhibit a maximum at some value of the magnetic field $B_{\textrm{max}}\sim B_{\textrm{c}}$ and then change sign with further increase in the value of the magnetic field. At fields much larger than the critical field, $B\gg B_{\textrm{c}}$, the second harmonic should approach zero. According to our discussion above, the value of the second harmonic at $B_{\textrm{max}}$ as well as the width of the peak must also depend on the disorder scattering rate. 
Thus, our goal in what follows will be to compute the second harmonic contribution to the current density as a function of the magnetic field and then determine the dependence of $B_{\textrm{max}}$ on model parameters and the frequency of the \emph{ac}-electric field. 

\section{Basic equations}
In this Section, we provide the basic information which will be required for the discussion of the nonlinear transport. 
\subsection{Model Hamiltonian}
In order to model the electron's system at the interface, we adopt the following Hamiltonian which describes non-interacting electrons which are constrained to move in a plane and are subjected to an external alternating vector potential and a static in-plane magnetic field ${\mathbf B}$: 
\beg\label{Eq1}
\begin{split}
\hat{\cal H}&=\frac{\hat{\mathbf P}^2}{2m}+\alpha_{\textrm{so}}\left({\vec e}_z\times\hat{\mbox{\boldmath $\sigma$}}\right)\cdot\hat{\mathbf P}+g\mu_B{\mathbf B}\hat{\mbox{\boldmath $\sigma$}}+U(\br).
\end{split}
\en
Here $\hat{\mathbf P}=\hat{\bp}-\frac{e}{c}{\mathbf A}(\br,t)$, $\hat{\bp}=-i{\mbox{\boldmath $\nabla$}}$ is electron momentum, $m$ is electron mass, $\alpha_{\textrm{so}}$ is the Rashba spin-orbit coupling constant, ${\mathbf B}=(B_x,B_y)$ is an external in-plane magnetic field, $g$ is the gyromagnetic factor, $U(\br)$ is a disorder potential and ${\mathbf A}(\br,t)$ is a vector potential of the periodically modulated electric field ${\mathbf E}=-(1/c)\partial_t{\mathbf A}={\mathbf E}_0e^{i(\bk\br-\omega t)}+c.c$.  In what follows we will assume for simplicity that the light is linearly polarized. We will assume that disorder is described by a correlation function
\beg\label{CorrDis}
\left\langle U(\br)U(\br')\right\rangle_{\textrm{dis}}=\frac{\delta(\br-\br')}{2\pi\nu_F\tau}.
\en
Here $\tau^{-1}$ is the disorder scattering rate and $\nu_F=m/2\pi$ is the single-particle density of states (per spin). 

\subsection{Kinetic equation for the Wigner distribution function}
The main quantity in our analysis below is the Wigner distribution function (WDF) for the electrons which is a $2\times 2$ matrix in spin space and is defined as follows
\beg\label{WDF}
\begin{aligned}
&w_{\alpha\beta}(\bp,\eps;\br,t)=\frac{1}{2\pi}\int d^2{\mathbf s}\int d\tau e^{i\bp{\mathbf s}-i\eps\tau}\\&\times\left\langle\psi_\beta\dg\left(\br+\frac{\mathbf s}{2},t+\frac{\tau}{2}\right)\psi_\alpha\left(\br-\frac{\mathbf s}{2},t-\frac{\tau}{2}\right)\right\rangle.
\end{aligned}
\en
Here $\psi_\alpha\dg(\br,t)$ and $\psi_\beta(\br,t)$ are usual fermionic creation and annihilation operators and $\alpha$ is the spin projection
on the $z$-axis.

In equilibrium and in the absence of the spin-orbit coupling ($\alpha_{\textrm{so}}=0$), for the WDF one finds
$\tilde{w}_{\alpha\beta}^{(0)}(\bk,\eps)=f(\eps)\delta(\eps-\eps_k)\delta_{\alpha\beta}$,
where $\eps_k$ is the single particle spectrum, $f(\eps)=\{\exp[(\eps-\mu)/k_BT]+1\}^{-1}$ is the Fermi-Dirac distribution function, and $\mu$ is the chemical potential. Consequently, in the presence of the spin-orbit coupling the WDF acquires the off-diagonal spin components which account for the possibility of having momentum pointing either along or opposite to the direction of the electron's spin. For this reason, it proves to be convenient to choose the basis in which the unperturbed WDF is diagonal. This basis is given by the eigenfunctions of the Hamiltonian (\ref{Eq1}) in which the vector and disorder potentials have been set to zero. The corresponding eigenvectors which define the chiral basis are given by 
\beg\label{Eigenvectors}
|u_{\bp 1}\rangle=\frac{1}{\sqrt{2}}\left(\begin{matrix} 1 \\ -e^{i\phi_{\mathbf b}} \end{matrix}\right), \quad
|u_{\bp 2}\rangle=\frac{1}{\sqrt{2}}\left(\begin{matrix} 1 \\ e^{i\phi_{\mathbf b}} \end{matrix}\right).
\en
Here we introduced $\phi_{\mathbf b}=\tan^{-1}({b_\bp^y}/{b_\bp^x})$ and 
\beg\label{Definebp}
{\mathbf b}_\bp=\alpha_{\textrm{so}}(\bp\times{\vec e}_z)+g\mu_B{\mathbf B}.  
\en
The angle $\phi_{\mathbf b}$ should not be confused with $\phi_\bp$
which defines the direction of the momentum $\bp=p(\cos\phi_\bp,
\sin\phi_\bp)$. In the basis of wave functions (\ref{Eigenvectors}) for the bare WDF, we have
\beg\label{NewBasis}
\hat{w}_{\bp\eps}^{(0)}=f(\eps)\left(\begin{matrix} \delta(\eps-\eps_{\bp 1}) & 0 \\ 0 & \delta(\eps-\eps_{\bp 2})
\end{matrix}\right)
\en
and $\eps_{\bp\alpha}$ are given by 
\beg\label{epsj}
\eps_{\bp\alpha}=\frac{\bp^2}{2m}+(-1)^\alpha b_\bp, ~(\alpha=1,2),
\en
representing the single particle dispersion of the two spin-orbit split bands. 

Following the avenue of Refs. \onlinecite{Andrey2006,Dzero:2024}, it can be shown that the WDF satisfies the following kinetic equation
\begin{equation}\label{Eq4w}
\begin{split}
&\left(\partial_t+\frac{1}{\tau}\right)\hat{w}_{\bp\eps}(t)+
i\left[{\mathbf b}_\bp\cdot{\mbox{\boldmath $\hat{\sigma}$}};\hat{w}_{\bp \eps}(t)\right]_{-}\\&=
-\frac{1}{2}\left[\frac{\bp}{m}+\alpha_{\textrm{so}}\hat{\mbox{\boldmath $\eta$}};\bm{\partial}\hat{w}_{\bp \eps}(t)\right]_++\St\{\hat{w}_{\bp\eps}\},
\end{split}
\end{equation}
where $[A;B]_{\pm}=AB\pm BA$, $\hat{\mbox{\boldmath $\eta$}}={\vec e}_z\times\hat{\mbox{\boldmath $\sigma$}}$, and we introduced the shorthand notation
\beg\label{TildeGrad}
\bm{\partial}\hat{w}_{\bp\eps}(t)=\frac{e}{\omega}\left[\hat{w}_{\bp\eps+\frac{\omega}{2}}(t)-\hat{w}_{\bp\eps-\frac{\omega}{2}}(t)\right]{\mathbf E}(t).
\en
The collision integral which appears in the right hand side of (\ref{Eq4w}) is due to disorder scattering and has the form 
\begin{equation}\label{Icoll}
\begin{split}
\St\{\hat{w}_{\bk\eps}\}=\frac{i}{2\pi\nu_{\text{F}}\tau}\int\frac{d^2\bk}{(2\pi)^2}
&\left\{\hat{G}_{\bk\eps}^R(t)\circ\hat{w}_{\bk\eps}(t)\right.\\&\left.-\hat{w}_{\bk\eps}(t)\circ\hat{G}_{\bk\eps}^A(t)\right\},
\end{split}
\en
where $\hat{G}_{\bk\eps}^{R(A)}(t)$ are the retarded (advanced) single particle propagators, and the convolution should be understood as 
$A_{\bk\veps}(t)\circ B_{\bk\veps}(t)=A_{\bk\veps}(t)e^{\frac{i}{2}(\stackrel{\leftarrow}\partial_\veps\stackrel{\rightarrow}\partial_t-\stackrel{\leftarrow}\partial_t\stackrel{\rightarrow}\partial_\veps)}B_{\bk\veps}(t)$.
After the kinetic equation for the WDF has been solved, the current density ${\mathbf j}(t)$ can be found using the standard expression
\beg\label{jtstandard}
{\mathbf j}(\br,t)=e\sum\limits_{\alpha,\beta=\up,\dn}\langle\psi_\alpha\dg(\br,t)\hat{\mathbf v}_{\alpha\beta}\psi_\beta(\br,t)\rangle,
\en
where $\hat{\mathbf v}=\partial\hat{H}/\partial \bp$ is the velocity operator. Employing the definitions (\ref{WDF},\ref{jtstandard}) we have
\beg\label{current}
{\mathbf j}(t)={e}\int\frac{d^2\bk}{(2\pi)^2}\int\limits_{-\infty}^{\infty}\textrm{Tr}\left\{\hat{\mathbf v}_{\bk}\hat{w}_{\bk\eps}(t)\right\}d\eps.
\en 
Note that this expression is general for it does not take into account that $\alpha_{\textrm{so}}\ll v_F$ ($v_F$ is the Fermi velocity).
The matrix elements of the velocity operator in the chiral basis (\ref{Eigenvectors}) are given by 
\beg\label{vxvyChiral}
\begin{split}
\hat{v}_x&=\left(\begin{matrix}
\frac{p_x}{m}+\alpha_{\textrm{so}}\sin\phi_{\mathbf b} & i\alpha_{\textrm{so}}\cos\phi_{\mathbf b} \\
-i\alpha_{\textrm{so}}\cos\phi_{\mathbf b} & \frac{p_x}{m}-\alpha_{\textrm{so}}\sin\phi_{\mathbf b}
\end{matrix}\right), \\
\hat{v}_y&=\left(\begin{matrix}
\frac{p_y}{m}-\alpha_{\textrm{so}}\cos\phi_{\mathbf b} & i\alpha_{\textrm{so}}\sin\phi_{\mathbf b} \\
-i\alpha_{\textrm{so}}\sin\phi_{\mathbf b} & \frac{p_y}{m}+\alpha_{\textrm{so}}\cos\phi_{\mathbf b}
\end{matrix}\right).
\end{split}
\en
The off-diagonal elements in both matrices are proportional to the components of the vector ${\mathbf b}_\bp$. In what follows, we will seek the perturbative solution of the kinetic equation in powers of the electric field. We will assume that the disorder is weak enough so that the contribution from the collision integral (\ref{Icoll}) can be fully ignored. Additionally we will adopt the limit of zero temperature, $T=0$.

\begin{figure}
\includegraphics[width=0.850\linewidth]{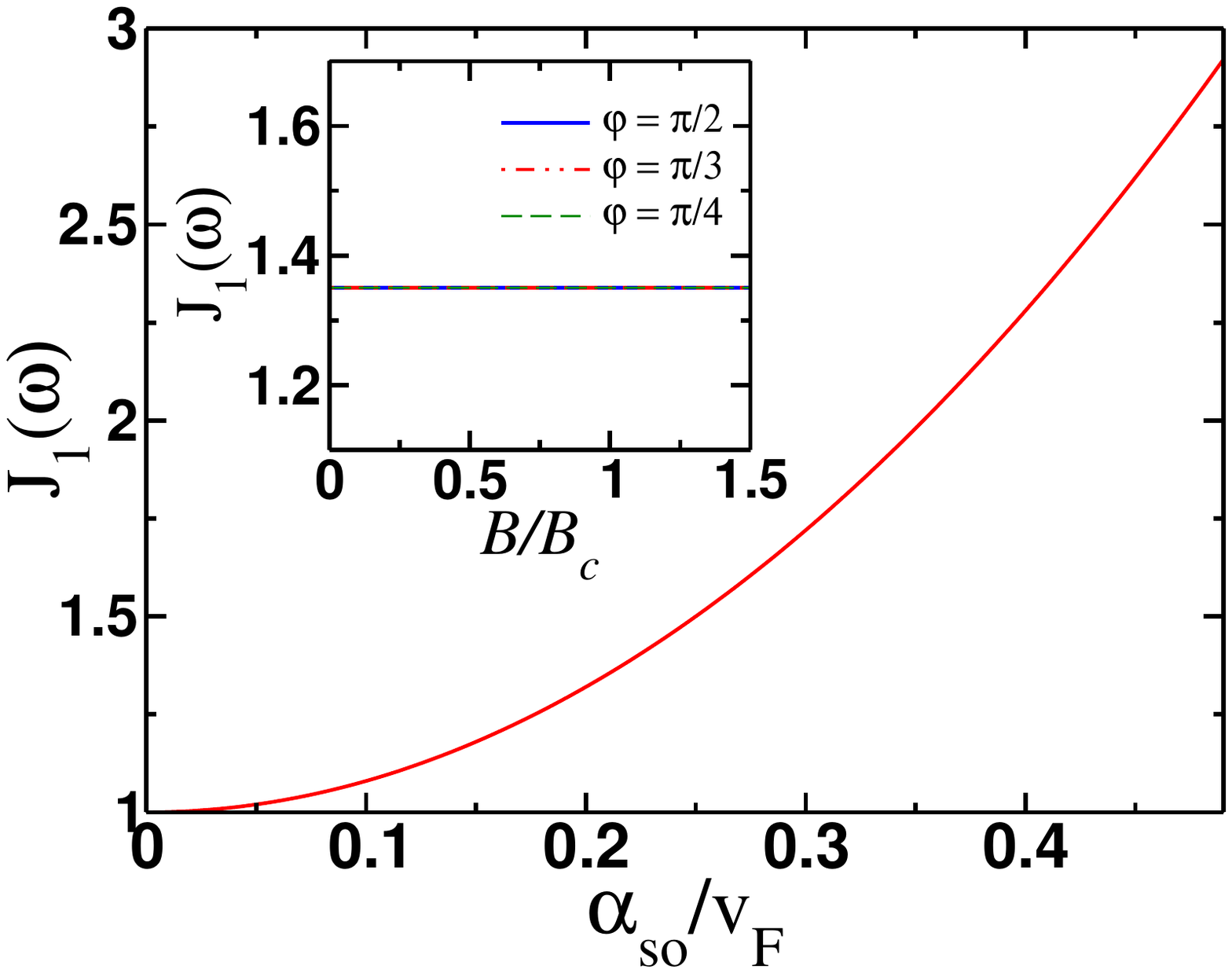} 
\caption{Dependence of the function $J_1(\omega)$, Eq. (\ref{LinearResponseFin}) on spin-orbit coupling (main panel) and on the magnetic field ${\mathbf B}=B(\cos\varphi,\sin\varphi)$ (inset). Here $B_c$ corresponds to the critical value of the magnetic field $g\mu_BB_c=\alpha_{\textrm{so}}p_F$ when the two spin-orbit split bands cross each other at the Dirac point.} 
\label{Fig1-j1w}
\end{figure}

\subsection{Linear response}
We start with the calculation of the first order in electric field correction to the current. We use the fact that by definition
matrix ${\mathbf b}_\bp\cdot{\mbox{\boldmath $\sigma$}}$ is diagonal in the chiral basis (\ref{Eigenvectors}):
\beg\label{SimplifyMatrix}
{\mathbf b}_\bp\cdot{\mbox{\boldmath $\sigma$}}=\left(\begin{matrix}-b_\bp & 0 \\ 0 & b_\bp\end{matrix}\right). 
\en
Then, the linear-in-electric-field corrections to the diagonal elements of the WDF can be easily found by evaluating the corresponding matrix elements using the basis set (\ref{Eigenvectors}) which yields
\beg\label{LinearDiag}
\begin{split}
\left[\hat{w}_{\bp\eps}^{(1)}\right]_{nn}=-\frac{e}{m}\left(\frac{\bp{\mathbf E}_0}{\omega z_\omega}\right)
\left[\hat{w}_{\bp\eps+\frac{\omega}{2}}^{(0)}-\hat{w}_{\bp\eps-\frac{\omega}{2}}^{(0)}\right]_{nn},
\end{split}
\en
where $n=1,2$ and $z_\omega=-i\omega+\tau^{-1}$. The term proportional to $\alpha_{\textrm{so}}(\hat{\mbox{\boldmath $\eta$}}\cdot{\mathbf E}_0)$ in the right hand side of equation (\ref{Eq4w}) does not contribute since its diagonal matrix elements are zero. The correction to the off-diagonal terms can be computed similarly to (\ref{LinearDiag}):
\beg\label{LinearOffDiag}
\begin{split}
\left[\hat{w}_{\bp\eps}^{(1)}\right]_{12}&=-\left(\frac{ie\alpha_{\textrm{so}}}{2\omega}\right)\frac{{\cal E}_x-i{\cal E}_y}{z_\omega-2ib_\bp}
\sum\limits_{n=1}^2\left[\hat{w}_{\bk\eps+\frac{\omega}{2}}^{(0)}-\hat{w}_{\bp\eps-\frac{\omega}{2}}^{(0)}\right]_{nn}, \\
\left[\hat{w}_{\bp\eps}^{(1)}\right]_{21}&=\left(\frac{ie\alpha_{\textrm{so}}}{2\omega}\right)\frac{{\cal E}_x+i{\cal E}_y}{z_\omega+2ib_\bp}
\sum\limits_{n=1}^2\left[\hat{w}_{\bp\eps+\frac{\omega}{2}}^{(0)}-\hat{w}_{\bp\eps-\frac{\omega}{2}}^{(0)}\right]_{nn}.
\end{split}
\en
Here we introduced ${\cal E}_\alpha={\mathbf E}_0\cdot{\vec e}_\alpha$ with $\alpha=x,y$.
With the help of expressions (\ref{vxvyChiral},\ref{LinearDiag},\ref{LinearOffDiag}), it is now straightforward to evaluate the linear response of the current density. We are primarily interested in computing the longitudinal contribution to the conductivity and for that reason we will set the $y$-component of the electric field to zero, ${\cal E}_y=0$. We refer the reader to the Appendix \ref{LinearResponse} for details of the calculation and here provide the final result:
\beg\label{LinearResponseFin}
\begin{aligned}
&j_x^{(\omega)}=
\left(\frac{e^2\nu_FD}{1-i\omega\tau}\right){\cal E}_x{J}_1(\omega).
\end{aligned}
\en
In this expression, $\nu_F=m/\pi$ is the single particle density of states at the Fermi level, $D=v_F^2\tau/2$ is the diffusion coefficient and function $J_{1}(\omega)$ have been defined in Appendix \ref{LinearResponse}.  Function ${J}_1(\omega)$ describes the intraband contribution (i.e. the contribution which involves the diagonal elements of the velocity matrix $\hat{v}_x$ only) and does not depend on the magnetic field while showing quadratic dependence on the spin-orbit coupling, Fig. \ref{Fig1-j1w}. Increase of the conductivity with the increase in the strength of spin-orbit coupling can be interpreted as being due to suppression of the momentum relaxation due to scattering on disorder. 
The fact that ${J}_1(\omega)$ is independent of the magnetic field can be easily understood by noting that the magnetic field leads to the relative change in the number of spin carriers, while the leading contribution to the conductivity must be proportional to the total number of carriers. 
In fact, the contribution which is governed by the interband scattering processes (i.e. the one which involves the off-diagonal matrix elements of $\hat{v}_x$) does indeed exhibit weak dependence on the magnetic field. However, it turns out to be proportional to $1/(\mu\tau)^2$  and, since $\mu\tau\gg 1$, it has been ignored. 

\begin{figure}
\includegraphics[width=0.850\linewidth]{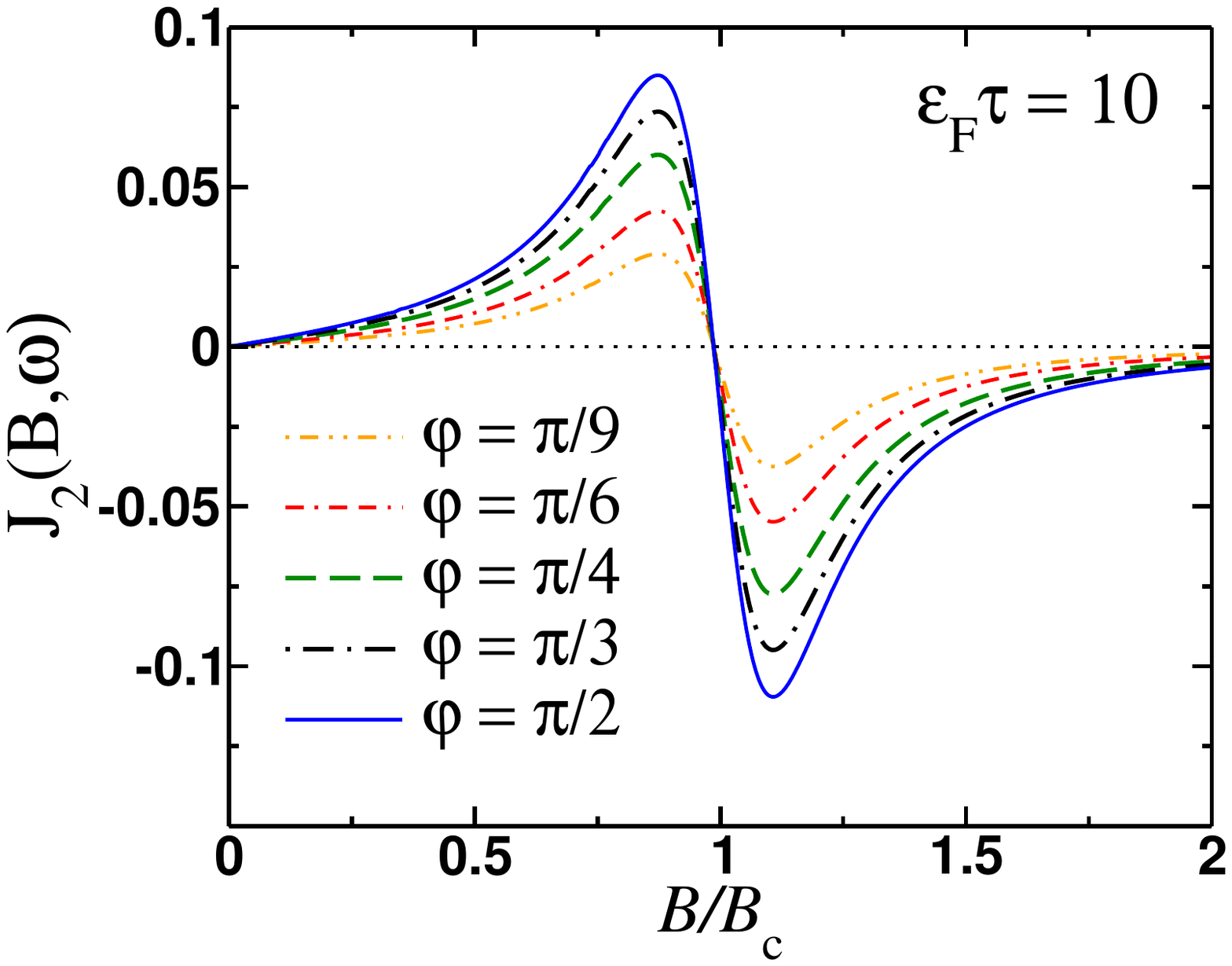} 
\caption{Dependence of the function $J_2({\mathbf B},\omega)$, Eq. (\ref{j2w}), on the magnetic field ${\mathbf B}=B(\cos\varphi,\sin\varphi)$. Here $B_{\textrm{c}}$ corresponds to the critical value of the magnetic field $g\mu_BB_{\textrm{c}}=\alpha_{\textrm{so}}p_F$ when the two spin-orbit split bands cross at the Dirac point. The positions of the maximum and minimum depend on the value of the dimensionless parameter $\veps_F\tau$. Note that the second harmonic changes sign at $B\approx B_{\textrm{c}}$.} 
\label{Fig2-j2w}
\end{figure}

\section{Second harmonic response}
An expression for the second harmonic can be obtained by solving (\ref{Eq4w}) to the second order in powers of the electric field. As it has already been mentioned, in the absence of a magnetic field and geometric (Berry) phase the second harmonic contribution to the current vanishes identically.
The calculation of the second harmonic is simplified by the following observation. In a two-dimensional electronic system with a band structure which has a nonzero Berry phase and lacks mirror symmetry, the second harmonic will be proportional to the Berry dipole.\cite{Deyo:2009,Sodemann:2015} Furthermore, within the formalism we use in this study, it can be shown that the Berry dipole contribution originates solely from the terms in (\ref{current}) which contain off-diagonal components of the velocity matrix. Since our model Hamiltonian possesses the mirror symmetry, these off-diagonal terms do not contribute to the second harmonic even for nonzero magnetic field, which can also be confirmed by a direct calculation. We find for the second harmonic the following expression [see Appendix B for details]:
\beg\label{j2w}
j_x^{(2\omega)}=-\frac{2m}{\pi}\left(\frac{e\alpha_{\textrm{so}}}{\Omega_{\textrm{so}}}\right)^3\left(\frac{1-i\omega\tau}{1-2i\omega\tau}\right)J_2({\mathbf B},\omega){\cal E}_x^2,
\en
where $\Omega_{\textrm{so}}=(\alpha_{\textrm{so}}p_F/\tau^2)^{1/3}$ and the dimensionless function $J_2({\mathbf B},\omega)$ is given by
\beg\label{J2Bw}
\begin{aligned}
J_2({\mathbf B},\omega)&=\frac{1}{2\pi}\left(\frac{\mu}{\omega}\right)^2\int\limits_{-\infty}^\infty\int\limits_{-\infty}^\infty
\frac{k_xdk_xdk_y}{(1-i\omega\tau)^2+4\tau^2b_\bk^2}\\&\times\sum\limits_{n=1}^2\left[\vartheta(\xi_{\bk n}+\omega)+\vartheta(\xi_{\bk n}-\omega)-2\vartheta(\xi_{\bk n})\right].
\end{aligned}
\en
Here $\xi_{\bk n}=\eps_{\bk n}-\mu$ and 
the pre-factor guarantees that this expression remains finite in the limit $\omega\to 0$. Note that in the limit of large frequencies $\omega\tau\gg 1$, the expression under the integral accounts for the resonant optical transitions between the two chiral bands.\cite{Dzero:2024} 

We start with the numerical analysis of the expression for the second harmonic. In Fig. \ref{Fig2-j2w} we show the dependence of this function on magnetic field assuming that the condition $\omega\tau\ll1$ holds. Quite expectedly, we find that the second harmonic response for the current along the $x$-axis is largest when the magnetic field is aligned along the $y$-axis, since quite generally one expects that ${\mathbf j}^{(2\omega)}$ to be proportional to the odd power of $({\vec e}_z\times{\mathbf B})$. \cite{Edelshtein:1988} As the magnitude of the magnetic field increases, the second harmonic response also increases and reaches the maximum value at fields $B\sim B_c$. This result seems to agree with the experimental observations by Tuvia et al. \onlinecite{Dagan:2024} who found a sharp increase in the values of $R_{2\omega}\propto\sigma_{2\omega}/\sigma_{1\omega}^2$ when increasing the magnetic field (we remind the reader that in our case $\sigma_{2\omega}\propto J_2({\mathbf B},\omega)$ and $\sigma_{1\omega}$ is the Drude conductivity multiplied by $J_1(\omega)$). It also must be mentioned that the position of the maximum (and minimum) as well as their widths depend on the value of the dimensionless parameter $\veps_F\tau$, Fig. \ref{Fig3-EFtau}. In particular, as the value of $\veps_F\tau$ approaches the 'dirty limit' $\veps_F\tau\sim1$ the peak of the second harmonic broadens and is reached for the magnetic field close to the critical value $g\mu_BB_c\approx\sqrt{2m\alpha_{\textrm{so}}^2\veps_F}$. This result implies that in the recent experiments \cite{Dagan:2024} the interfaces should be generally described in the diffusive limit $p_Fl\sim 1$, where $l$ is the mean-free path. 

Lastly, from our results shown in Fig. \ref{Fig2-j2w}  we note that the second harmonic changes sign when the magnetic field reaches the value $B_0$, and this effect is independent of the direction of the magnetic field.  It needs to be emphasized here that, as it follows from our numerical calculations, vanishing of the second harmonic is due to the opposite in sign but equal in magnitude contributions from the two chiral bands. It appears that in order to observe this effect experimentally one requires a fairly clean interface with $\veps_F\tau\sim 10$ and also higher magnetic fields than the ones reported earlier. \cite{Dagan:2024}

\begin{figure}
\includegraphics[width=0.850\linewidth]{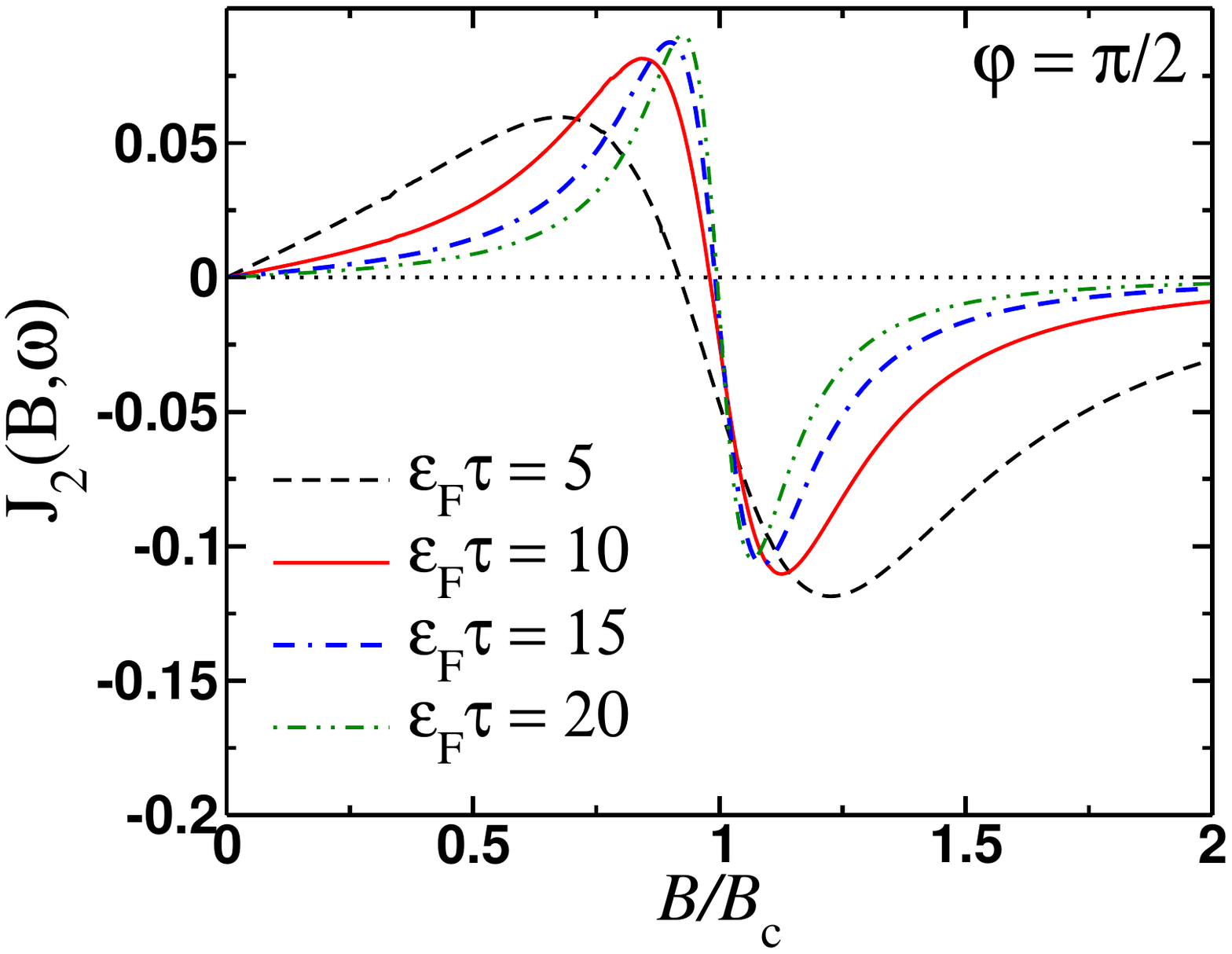} 
\caption{Dependence of the function $J_2({\mathbf B},\omega)$, Eq. (\ref{j2w}), on the magnitude of the in-plane magnetic field $B$ for various values of the dimensionless parameter $\veps_F\tau$, where $\veps_F$ is the Fermi energy and $\tau^{-1}$ is the disorder scattering rate (\ref{CorrDis}). The magnetic field was assumed to be applied along the $y$-axis.} 
\label{Fig3-EFtau}
\end{figure}

Let us now focus on obtaining the approximate values of the magnetic field when function $J_2({\mathbf B},\omega)$ reaches its maximum and minimum. This will also allow us to get an estimate for the value of the magnetic field when the second harmonic changes sign. 
In addition, without loss of generality we consider the case when the magnetic field points along the $y$-axis, ${\mathbf B}=B{\vec e}_y$. In the integrals (\ref{J2Bw}) we rescale $k_\alpha\to p_Fk_\alpha$, introduce the dimensionless frequency parameter $\overline{\omega}=\omega/\mu$, and use the polar coordinates. Assuming $\omega\tau\ll 1$, (\ref{J2Bw}) can be approximately re-written as follows
\beg\label{J2Bwk}
\begin{aligned}
J_2(B,\omega)&=\frac{p_F^3}{2\pi}\int\limits_{0}^{2\pi}{\cal I}_\omega(Q,\varphi)\cos\varphi{d\varphi},
\end{aligned}
\en
where $Q=g\mu_BB/(\alpha_{\textrm{so}}p_F)$ and
\beg\label{Iphi}
{\cal I}_\omega(Q,\varphi)=\frac{1}{\overline{\omega}^2}\sum\limits_{n=1}^2\left\{\int\limits_{k_{\omega n}^{-}}^{k_{0n}}
\frac{k^2 dk}{1+4\zeta^2r_k^2}
-\int\limits_{k_{0n}}^{k_{\omega n}^{+}}
\frac{k^2 dk}{1+4\zeta^2r_k^2}\right\}.
\en
Here $r_k^2=k^2+Q^2-2kQ\cos\varphi$, $\zeta=\tau\alpha_{\textrm{so}}p_F$, $k_{0n}(\varphi)$ denotes the positive root of the equation $\xi_{\bk n}=0$ and $k_{\omega n}^{\pm}(\varphi)$ similarly denote the positive roots of the equations $\xi_{\bk n}=\pm {\omega}$ correspondingly (note that $k_{0n}$ and $k_{\omega n}^{\pm}$ are taken in the units of the Fermi momentum, $p_F$). The critical value of the magnetic field for which the bands cross at the chemical potential corresponds to $Q_\textrm{cr.}=1$. 

Our numerical analysis of the function ${\cal I}_\omega(Q,\varphi)$ shows that for $Q^*\approx Q_{\textrm{cr.}}$ one finds that ${\cal I}_\omega(Q^*,\varphi)={\cal I}_\omega(Q^*)$, i.e. it remains independent of $\varphi$. Therefore, in expression (\ref{J2Bwk}) upon integration over $\varphi$, the second harmonic vanishes. Whether $Q^{*}$ equals exactly $Q_{\textrm{cr.}}$ or not depends on the value of the dimensionless parameter $\veps_F\tau$, Fig. \ref{Fig3-EFtau}. 

To get further insight into the angular dependence of function ${\cal I}_\omega(Q,\varphi)$, we take into account that $\overline{\omega}\ll 1$. Given that $k_{0n}(\varphi)\sim 1$, this allows one to expand $k_{\omega n}^{\pm}(\varphi)$ up to the second order in $\overline{\omega}$ followed up by expanding the integrals (\ref{Iphi}) also up to the second order in powers of $\overline{\omega}$. For example, for the case $\varphi=\pi$ we find
\beg\label{phipi}
\begin{split}
{\cal I}_\omega(Q_{\textrm{cr}},\pi)&\approx\sum\limits_{n=1}^2\left[\frac{2\beta_{n}}{1+4\zeta^2(q_{0n}+1)^2}\right.\\&\left.-\frac{2\tilde{\beta}_n^2q_{0n}[1+4\zeta^2(q_{0n}+1)]}{[1+4\zeta^2(q_{0n}+1)^2]^2}\right]\approx 0,
\end{split}
\en
where $q_{0n}=1-2\lambda_n$, $\lambda_n=(-1)^2(\alpha_{\textrm{so}}/v_F)$, $\beta_n=[2(1-\lambda_n)]^{-3}$ and $\tilde{\beta}_n=[2(1-\lambda_n)]^{-1}$.
The same result can be found for other values of $\varphi$ which confirms that ${\cal I}_\omega(Q_{\textrm{cr}},\varphi)={\cal I}_\omega(Q_{\textrm{cr}})$. Lastly, we note that the value of the magnetic field for which the second harmonic reaches its maximum value has to be determined analytically since it strongly depends on $\veps_F\tau$. From our estimates it follows that for $\veps_F\tau\gg 1$ the maximum value is reached for $Q\sim1$, i.e. when $g\mu_BB_c\approx\alpha p_F$.

\section{Summary}
In this work, we have considered nonlinear transport in a two-dimensional disordered electron gas in the presence of strong Rashba spin-orbit coupling and under an externally applied in-plane magnetic field. Our theory uses a quantum kinetic equation which in principle allows one to also explore the transport in $ac$-regime at arbitrary frequencies. We found that the nonlinear resistance changes sign as a function of the magnetic field and vanishes in the limits of very strong fields. In contrast to the previous work, \cite{Dagan:2024} we find that the second harmonic resistance does not vanish at fields above the critical one but changes sign. This change reflects the position of the Dirac point relative to the Fermi contours of the spin-orbit split band structure. The value of the magnetic field where the second harmonic changes sign shows a weak dependence on the disorder scattering rate. 

\section{Acknowledgments}
We thank Alex Levchenko for useful discussions on various aspects of this study. This work was financially supported by the National Science Foundation grant DMR-2400484 (A.S. and M.D.) M. D. has performed this work in part at Aspen Center for Physics, which is supported by the National Science Foundation grant PHY-2210452. M. K. acknowledges the support of the grant NSF-BSF DMR-2023693.

\begin{appendix}
\section{Linear response}\label{LinearResponse}
In this section we provide the details of the calculation of the linear response current (\ref{LinearResponseFin}). We start by considering the contributions from the diagonal elements  of the WDF:
\beg\label{J1dia}
\begin{split}
&[\hat{v}_x]_{11}\left[\hat{w}_{\bk\eps}^{(1)}\right]_{11}+[\hat{v}_x]_{22}\left[\hat{w}_{\bk\eps}^{(1)}\right]_{22}=\\&-\left(\frac{ek_x}{m^2\omega}\right)\left(\frac{\bk{\mathbf E}}{z_\omega}\right)
\sum\limits_{n=1}^2\left[\hat{w}_{\bk\eps+\frac{\omega}{2}}^{(0)}-\hat{w}_{\bk\eps-\frac{\omega}{2}}^{(0)}\right]_{nn}\\&+\left(\frac{e\alpha_{\textrm{so}}}{m\omega}\right)\left(\frac{\bk{\mathbf E}}{z_\omega}\right)\left(\frac{b_\bk^y}{b_\bk}\right)\sum\limits_{n=1}^2(-1)^n\left[\hat{w}_{\bk\eps+\frac{\omega}{2}}^{(0)}-\hat{w}_{\bk\eps-\frac{\omega}{2}}^{(0)}\right]_{nn}.
\end{split}
\en
In the subsequent integration over momentum we can re-scale all momenta in terms of the Fermi momentum. As a result the second term on the right hand side turns out to be of the order of $\alpha_{\textrm{so}}/v_F$ smaller compared to the first one. In addition, these terms contain another small pre-factor due to the kinematic constraints and therefore it will be ignored. 

Subsequent integration of the first term in the right hand side (\ref{J1dia}) over $\eps$ is trivial due to definition (\ref{NewBasis}). We consider function
\beg\nonumber
\begin{split}{J}_{1}({\mathbf B},\omega)&=\frac{1}{p_F^4}\sum\limits_{s=\pm}\int\limits_{-\infty}^\infty\int\limits_{-\infty}^\infty
\frac{k_x^2dk_xdk_y}{2\pi}\left[\vartheta\left(\mu-\frac{k^2}{2m}+\omega-sb_\bk\right)\right.\\&\left.-\vartheta\left(\mu-\frac{k^2}{2m}-\omega-sb_\bk\right)\right].
\end{split}
\en
In order to evaluate this integral, one may introduce the new momentum $\alpha_{\textrm{so}}{\bq}=\alpha_{\textrm{so}}{\bk}\times{\vec e}_z+g\mu_B{\mathbf B}$, which after using the polar coordinates in principle allows one to evaluate an integral over $q$ analytically. Alternatively, one can compute the limits of integration on $q$ by resolving the step functions numerically and then integrate over $q$ analytically followed up by the numerical integration over the polar angle. We find that the function ${J}_{1}({\mathbf B},\omega)$ is in fact independent of ${\mathbf B}$ and increases with increasing value of the spin-orbit coupling, ${J}_{1}({\mathbf B},\omega)\propto\alpha_{\textrm{so}}^2$.

\section{Nonlinear response}
\subsection{Contribution from the diagonal components of the velocity matrix}
As we have discussed in the main text, the leading contribution to the second harmonic originates from the terms which contain the diagonal elements of the velocity matrix. Expressions for the second order corrections to the diagonal elements of $\hat{w}_{\bk\eps}^{(2)}$ are
\beg\label{w2nn}
\begin{aligned}
[\hat{w}_{\bk\eps}^{(2)}]_{\alpha\alpha}&=-\frac{e}{m}\left(\frac{\bk{\mathbf E}}{\omega z_{2\omega}}\right)
\left[\hat{w}_{\bk\eps+\frac{\omega}{2}}^{(1)}-\hat{w}_{\bk\eps-\frac{\omega}{2}}^{(1)}\right]_{\alpha\alpha}\\
&-\left(\frac{e\alpha_{\textrm{so}}}{2}\right)\frac{\left(\hat{\mbox{\boldmath $\eta$}}{\mathbf E}\right)_{\alpha\overline{\alpha}}}{\omega z_{2\omega}}
\left[\hat{w}_{\bk\eps+\frac{\omega}{2}}^{(1)}-\hat{w}_{\bk\eps-\frac{\omega}{2}}^{(1)}\right]_{\overline{\alpha}\alpha}\\
&-\left(\frac{e\alpha_{\textrm{so}}}{2}\right)\frac{\left(\hat{\mbox{\boldmath $\eta$}}{\mathbf E}\right)_{\overline{\alpha}\alpha}}{\omega z_{2\omega}}
\left[\hat{w}_{\bk\eps+\frac{\omega}{2}}^{(1)}-\hat{w}_{\bk\eps-\frac{\omega}{2}}^{(1)}\right]_{\alpha\overline{\alpha}}, \\
\end{aligned}
\en
Here $z_{2\omega}=-2i\omega+{\tau}^{-1}$. Using the expressions (\ref{LinearDiag},\ref{LinearOffDiag}) it is straightforward to find the expression for the second harmonic due to the diagonal components of the velocity matrix:
\beg\label{DiaSum}
\begin{split}
&\int\limits_{-\infty}^\infty\left([\hat{v}_x]_{11}\left[\hat{w}_{\bk\eps}^{(2)}\right]_{11}+[\hat{v}_x]_{22}\left[\hat{w}_{\bk\eps}^{(2)}\right]_{22}\right)d\eps\\&=\frac{e^2k_x{\cal E}_x^2}{m\omega^2z_\omega z_{2\omega}}\left\{
\left(\frac{k_x^2}{m}
+\frac{\alpha_{\textrm{so}}^2z_\omega^2}{(z_\omega^2+4b_\bk^2)}\right)
\sum\limits_{n=1}^2{\cal W}(\eps_{\bk n},\omega)\right.\\&\left.+
{\alpha_{\textrm{so}}\left(\frac{k_x}{m}\right)}\left(\frac{b_\bk^y}{b_\bk}\right)\sum\limits_{n=1}^2(-1)^n{\cal W}(\eps_{\bk n},\omega)
\right\},
\end{split}
\en
where function ${\cal W}(\eps_{\bk n},\omega)$ is defined according to
\beg\label{Wkn}
\begin{split}
{\cal W}(\eps_{\bk n},\omega)&=\vartheta\left(\mu+\omega-\eps_{\bk n}\right)+\vartheta\left(\mu-\omega-\eps_{\bk n}\right)\\&-2\vartheta\left(\mu-\eps_{\bk n}\right),
\end{split}
\en
where we again employed the limit of low temperatures. Note that the first and the last term on the right hand side (\ref{DiaSum}) are similar to the terms which appear in the expression (\ref{J1dia}) for the current density in the linear response. 

It will be beneficial to analyze each of the three terms (\ref{DiaSum}) separately. Let us start with the first one whose contribution is determined by the following integral
\beg\label{INt1}
\int\frac{d^2\bk}{(2\pi)^2}\left(\frac{k_x}{m}\right)^3\sum\limits_{n=1}^2{\cal W}(\eps_{\bk n},\omega).
\en
Clearly, this integral vanishes identically when ${\mathbf B}=0$. Upon closer inspection one also finds that this integral must be an odd function of the magnetic field. This in turn implies that upon expanding the expression under the integrals in powers of the magnetic field  there will appear an overall pre-factor $(-1)^n$ which renders the whole expression to become zero upon summation over $n$. 

Next, we consider the third term whose contribution to the second harmonic is determined by the following integral
\beg\label{INt3}
\alpha_{\textrm{so}}\int\frac{d^2\bk}{(2\pi)^2}\left(\frac{b_\bk^y}{b_\bk}\right)\left(\frac{k_x}{m}\right)^2\sum\limits_{n=1}^2(-1)^n{\cal W}(\eps_{\bk n},\omega)
\en
with $b_\bk^y=-\alpha_{\textrm{so}}k_x+g\mu_BB_y$, Eq. (\ref{Definebp}). It can be also shown that this integral vanishes both numerically and analytically. In the analytic calculation one needs to use the new integration variables which correspond to the shifted momentum 
$\alpha_{\textrm{so}}\bq=\alpha_{\textrm{so}}\left(\bk\times{\vec e}_z\right)+g\mu_B{\mathbf B}$. Expressed in terms of these new variables, $b_\bk\to \alpha_{\textrm{so}}q$. Then upon the series expansion of the expression under the integral in powers of ${\mathbf B}$, it follows that the terms which contain the odd powers of ${\mathbf B}$ are vanishingly small. Thus, the remaining contribution is of the form
\beg\label{SecondHarm}
\begin{split}
j_x^{(2\omega)}&=\frac{e^3\alpha_{\textrm{so}}^2{\cal E}_x^2}{4\pi^2m\omega^2}\left(\frac{1-i\omega\tau}{1-2i\omega\tau}\right)\int\limits_{-\infty}^\infty\int\limits_{-\infty}^\infty
\frac{k_xdk_xdk_y}{z_\omega^2+4b_\bk^2}\\&\times\sum\limits_{n=1}^2\left[f(\eps_{\bk n}+\omega)+f(\eps_{\bk n}-\omega)-2f(\eps_{\bk n})\right]
\end{split}
\en
and $z_{\omega}=-i\omega+\tau^{-1}$. One readily sees that in the limit $B\to 0$ this integral vanishes identically. In the limit of small fields the integral will be proportional to $B_y$: this result can be found by expanding the function $b_\bk$ to the linear order in powers of 
${\mathbf B}$. For further discussion it is convenient to re-write (\ref{SecondHarm}) in the form given by (\ref{j2w}) in the main text. 
\subsection{Contribution from the off-diagonal components of the velocity matrix}
We now briefly turn our attention to the calculation of the second harmonic from the off-diagonal components of the velocity matrix. We have
\beg\label{j2woff}
\begin{split}
&\left[{j}_x^{(2\omega)}\right]_{\textrm{off-diag.}}={e}\int_{\bk,\eps}\left\{[\hat{v}_x]_{12}\left[\hat{w}_{\bk\eps}^{(2)}\right]_{21}+[\hat{v}_x]_{21}\left[\hat{w}_{\bk\eps}^{(2)}\right]_{12}\right\}\\&=ie\alpha_{\textrm{so}}
\int\frac{d^2\bk}{(2\pi)^2}\left(\frac{b_\bk^x}{b_\bk}\right)\int\limits_{-\infty}^{\infty}\left\{\left[\hat{w}_{\bk\eps}^{(2)}\right]_{21}-\left[\hat{w}_{\bk\eps}^{(2)}\right]_{12}\right\}d\eps
\end{split}
\en
and on the second step we used (\ref{vxvyChiral}).
The second order corrections to the off-diagonal elements of the WDF are given by 
\begin{widetext}
\beg\label{w212}
\begin{split}
\left[\hat{w}_{\bk\eps}^{(2)}\right]_{12}=&
-\left(\frac{e{\cal E}_x}{2\omega}\right)\left(\frac{i\alpha_{\textrm{so}}}{z_{2\omega}-2ib_\bk}\right)
\sum\limits_{n=1}^2\left[\hat{w}_{\bk\eps+\frac{\omega}{2}}^{(1)}-\hat{w}_{\bk\eps-\frac{\omega}{2}}^{(1)}\right]_{nn}-\left(\frac{e{\cal E}_x}{\omega}\right)\left(\frac{k_x}{m}\right)\frac{\left[\hat{w}_{\bk\eps+\frac{\omega}{2}}^{(1)}-\hat{w}_{\bk\eps-\frac{\omega}{2}}^{(1)}\right]_{12}}{z_{2\omega}-2ib_\bk}
, \\
\left[\hat{w}_{\bk\eps}^{(2)}\right]_{21}=&\left(\frac{e{\cal E}_x}{2\omega}\right)\left(\frac{i\alpha_{\textrm{so}}}{z_{2\omega}+2ib_\bk}\right)
\sum\limits_{n=1}^2\left[\hat{w}_{\bk\eps+\frac{\omega}{2}}^{(1)}-\hat{w}_{\bk\eps-\frac{\omega}{2}}^{(1)}\right]_{nn}
-\left(\frac{e{\cal E}_x}{\omega}\right)\left(\frac{k_x}{m}\right)\frac{\left[\hat{w}_{\bk\eps+\frac{\omega}{2}}^{(1)}-\hat{w}_{\bk\eps-\frac{\omega}{2}}^{(1)}\right]_{21}}{z_{2\omega}+2ib_\bk}.
\end{split}
\en
\end{widetext}
Here we took into account that the electric field points along the $x$-axis. We finally proceed with discussing two of these contributions separately. For the first one we find
\beg\label{j2woffa}
\begin{split}
&\left[{j}_{x,\textrm{1}}^{(2\omega)}({\mathbf B})\right]_{\textrm{off-diag.}}=\left(\frac{e^3\alpha_{\textrm{so}}^2}{\omega^2}\right)\left(\frac{{\cal E}^2}{z_\omega z_{2\omega}}\right)\int\frac{d^2{\bk}}{(2\pi)^2}\left(\frac{k_x}{m}\right)\\&\times\left(\frac{b_\bk^x}{b_\bk}\right)\frac{z_{2\omega}^2}{z_{2\omega}^2+4b_\bk^2}\sum\limits_{n=1}^2{\cal W}(\eps_{\bk n},\omega).
\end{split}
\en
Expectedly, this expression vanishes when ${\mathbf B}=0$. It also vanishes if we consider the linear-in-magnetic field corrections since $b_\bk^x=\alpha_{\textrm{so}}k_y+g\mu_BB_x$. Calculation of the momentum integrals can be done similarly to how it was done above. From numerics we also find that (\ref{j2woffa}) vanishes identically.  The same applies to the remaining contribution. Thus, we conclude that the off-diagonal components of the velocity do not contribute to the second harmonic unless there is nonzero geometric phase. 

\end{appendix}


\begin{thebibliography}{10}

\bibitem{OxideInterfaces-Review}
P.~Yu, Y.-H. Chu, and R.~Ramesh, ``Oxide interfaces: pathways to novel
  phenomena,'' {\em Materials Today}, vol.~15, no.~7, pp.~320--327, 2012.

\bibitem{OxideInterfaces-2023}
K.~Ravindran, J.~K. Dey, A.~Keshri, B.~Roul, S.~B. Krupanidhi, and S.~Das,
  ``Topological phenomena at the oxide interfaces,'' {\em Materials for Quantum
  Technology}, vol.~3, p.~012002, mar 2023.

\bibitem{SrTi03}
A.~F. Santander-Syro, F.~Fortuna, C.~Bareille, T.~C. R{\"o}del, G.~Landolt,
  N.~C. Plumb, J.~H. Dil, and M.~Radovi{\'c}, ``Giant spin splitting of the
  two-dimensional electron gas at the surface of $\mathrm{SrTiO}_3$,'' {\em Nature
  Materials}, vol.~13, no.~12, pp.~1085--1090, 2014.

\bibitem{SOCoupling}
S.~LaShell, B.~A. McDougall, and E.~Jensen, ``Spin splitting of an $\mathrm{Au}$(111)
  surface state band observed with angle resolved photoelectron spectroscopy,''
  {\em Phys. Rev. Lett.}, vol.~77, pp.~3419--3422, Oct 1996.

\bibitem{Rashba-SO}
A.~Manchon, H.~C. Koo, J.~Nitta, S.~M. Frolov, and R.~A. Duine, ``New
  perspectives for Rashba spin--orbit coupling,'' {\em Nature Materials},
  vol.~14, no.~9, pp.~871--882, 2015.

\bibitem{BMER-2018}
P.~He, S.~M. Walker, S.~S.-L. Zhang, F.~Y. Bruno, M.~S. Bahramy, J.~M. Lee,
  R.~Ramaswamy, K.~Cai, O.~Heinonen, G.~Vignale, F.~Baumberger, and H.~Yang,
  ``Observation of out-of-plane spin texture in a ${\mathrm{SrTiO}}_{3}(111)$
  two-dimensional electron gas,'' {\em Phys. Rev. Lett.}, vol.~120, p.~266802,
  Jun 2018.

\bibitem{Lesne-2023}
E.~Lesne, Y.~G. Sa{\v g}lam, R.~Battilomo, M.~T. Mercaldo, T.~C. van Thiel,
  U.~Filippozzi, C.~Noce, M.~Cuoco, G.~A. Steele, C.~Ortix, and A.~D. Caviglia,
  ``Designing spin and orbital sources of Berry curvature at oxide
  interfaces,'' {\em Nature Materials}, vol.~22, no.~5, pp.~576--582, 2023.

\bibitem{MR-2017}
P.~K. Rout, I.~Agireen, E.~Maniv, M.~Goldstein, and Y.~Dagan, ``Six-fold
  crystalline anisotropic magnetoresistance in the (111)
  ${\mathrm{LaAlO}}_{3}/{\mathrm{SrTiO}}_{3}$ oxide interface,'' {\em Phys.
  Rev. B}, vol.~95, p.~241107, Jun 2017.

\bibitem{Non-reciprocal2019}
D.~Choe, M.-J. Jin, S.-I. Kim, H.-J. Choi, J.~Jo, I.~Oh, J.~Park, H.~Jin, H.~C.
  Koo, B.-C. Min, S.~Hong, H.-W. Lee, S.-H. Baek, and J.-W. Yoo, ``Gate-tunable
  giant nonreciprocal charge transport in noncentrosymmetric oxide
  interfaces,'' {\em Nature Communications}, vol.~10, no.~1, p.~4510, 2019.

\bibitem{Dagan-2010SC}
M.~Ben~Shalom, M.~Sachs, D.~Rakhmilevitch, A.~Palevski, and Y.~Dagan, ``Tuning
  spin-orbit coupling and superconductivity at the
  ${\mathrm{SrTiO}}_{3}/{\mathrm{LaAlO}}_{3}$ interface: A magnetotransport
  study,'' {\em Phys. Rev. Lett.}, vol.~104, p.~126802, Mar 2010.

\bibitem{Dagan-2010}
D.~Rakhmilevitch, M.~Ben~Shalom, M.~Eshkol, A.~Tsukernik, A.~Palevski, and
  Y.~Dagan, ``Phase coherent transport in
  ${\mathrm{SrTiO}}_{3}/{\mathrm{LaAlO}}_{3}$ interfaces,'' {\em Phys. Rev. B},
  vol.~82, p.~235119, Dec 2010.

\bibitem{Dagan-Link2017}
P.~K. Rout, E.~Maniv, and Y.~Dagan, ``Link between the superconducting dome and
  spin-orbit interaction in the (111)
  ${\mathrm{LaAlO}}_{3}/\mathrm{SrTi}{\mathrm{O}}_{3}$ interface,'' {\em Phys.
  Rev. Lett.}, vol.~119, p.~237002, Dec 2017.

\bibitem{Altman2012}
A.~Joshua, S.~Pecker, J.~Ruhman, E.~Altman, and S.~Ilani, ``A universal
  critical density underlying the physics of electrons at the $\mathrm{LaAlO}_3$/$\mathrm{SrTiO}_3$
  interface,'' {\em Nature Communications}, vol.~3, no.~1, p.~1129, 2012.

\bibitem{Altman2013}
A.~Joshua, J.~Ruhman, S.~Pecker, E.~Altman, and S.~Ilani, ``Gate-tunable
  polarized phase of two-dimensional electrons at the
  $\mathrm{LaAlO}_3$$\mathrm{SrTiO}_3$ interface,'' {\em Proceedings of the
  National Academy of Sciences}, vol.~110, no.~24, pp.~9633--9638, 2013.

\bibitem{Biscaras2010}
J.~Biscaras, N.~Bergeal, A.~Kushwaha, T.~Wolf, A.~Rastogi, R.~C. Budhani, and
  J.~Lesueur, ``Two-dimensional superconductivity at a mott insulator/band
  insulator interface $\mathrm{LaTiO}_3$/$\mathrm{SrTiO}_3$,'' {\em Nature Communications}, vol.~1,
  no.~1, p.~89, 2010.

\bibitem{SH2017}
T.~Ideue, K.~Hamamoto, S.~Koshikawa, M.~Ezawa, S.~Shimizu, Y.~Kaneko,
  Y.~Tokura, N.~Nagaosa, and Y.~Iwasa, ``Bulk rectification effect in a polar
  semiconductor,'' {\em Nature Physics}, vol.~13, no.~6, pp.~578--583, 2017.

\bibitem{Rectify1}
H.~Isobe, S.-Y. Xu, and L.~Fu, ``High-frequency rectification via chiral Bloch
  electrons,'' {\em Science Advances}, vol.~6, no.~13, p.~eaay2497, 2020.

\bibitem{FDoubling2021}
P.~He, H.~Isobe, D.~Zhu, C.-H. Hsu, L.~Fu, and H.~Yang, ``Quantum frequency
  doubling in the topological insulator $\mathrm{Bi}_2\mathrm{Se}_3$,'' {\em Nature Communications},
  vol.~12, no.~1, p.~698, 2021.

\bibitem{NR-Transport}
Y.~M. Itahashi, T.~Ideue, Y.~Saito, S.~Shimizu, T.~Ouchi, T.~Nojima, and
  Y.~Iwasa, ``Nonreciprocal transport in gate-induced polar superconductor
  $\mathrm{SrTiO}_3$,'' {\em Science Advances}, vol.~6, no.~13, p.~eaay9120,
  2020.

\bibitem{NR-Transport2}
Y.~M. Itahashi, T.~Ideue, S.~Hoshino, C.~Goto, H.~Namiki, T.~Sasagawa, and
  Y.~Iwasa, ``Giant second harmonic transport under time-reversal symmetry in a
  trigonal superconductor,'' {\em Nature Communications}, vol.~13, no.~1,
  p.~1659, 2022.

\bibitem{PlanarHall2021}
R.~Battilomo, N.~Scopigno, and C.~Ortix, ``Anomalous planar Hall effect in
  two-dimensional trigonal crystals,'' {\em Phys. Rev. Res.}, vol.~3,
  p.~L012006, Jan 2021.

\bibitem{NonlinearMT2019}
P.~He, C.-H. Hsu, S.~Shi, K.~Cai, J.~Wang, Q.~Wang, G.~Eda, H.~Lin, V.~M.
  Pereira, and H.~Yang, ``Nonlinear magnetotransport shaped by Fermi surface
  topology and convexity,'' {\em Nature Communications}, vol.~10, no.~1,
  p.~1290, 2019.

\bibitem{CIMOKE}
E.~J. K\"onig, M.~Dzero, A.~Levchenko, and D.~A. Pesin, ``Gyrotropic Hall
  effect in Berry-curved materials,'' {\em Phys. Rev. B}, vol.~99, p.~155404,
  Apr 2019.

\bibitem{Spivak2009}
E.~Deyo, L.~E. Golub, E.~L. Ivchenko, and B.~Spivak, ``Semiclassical theory of
  the photogalvanic effect in non-centrosymmetric systems,'' arXiv:0904.1917
  (2009).

\bibitem{Sodemann2015}
I.~Sodemann and L.~Fu, ``Quantum nonlinear Hall effect induced by Berry
  curvature dipole in time-reversal invariant materials,'' {\em Phys. Rev.
  Lett.}, vol.~115, p.~216806, Nov 2015.

\bibitem{BerryDipole1}
T.~Low, Y.~Jiang, and F.~Guinea, ``Topological currents in black phosphorus
  with broken inversion symmetry,'' {\em Phys. Rev. B}, vol.~92, p.~235447, Dec
  2015.

\bibitem{BerryDipole2}
J.~I. Facio, D.~Efremov, K.~Koepernik, J.-S. You, I.~Sodemann, and J.~van~den
  Brink, ``Strongly enhanced Berry dipole at topological phase transitions in
  $\mathrm{BiTeI}$,'' {\em Phys. Rev. Lett.}, vol.~121, p.~246403, Dec 2018.

\bibitem{BerryDipole3}
J.-S. You, S.~Fang, S.-Y. Xu, E.~Kaxiras, and T.~Low, ``Berry curvature dipole
  current in the transition metal dichalcogenides family,'' {\em Phys. Rev. B},
  vol.~98, p.~121109, Sep 2018.

\bibitem{BerryDipole4}
Y.~Zhang, Y.~Sun, and B.~Yan, ``Berry curvature dipole in Weyl semimetal
  materials: an ab initio study,'' {\em Phys. Rev. B}, vol.~97, p.~041101, Jan
  2018.

\bibitem{BerryDipole5}
Z.~Z. Du, C.~M. Wang, H.-Z. Lu, and X.~C. Xie, ``Band signatures for strong
  nonlinear Hall effect in bilayer ${\mathrm{WTe}}_{2}$,'' {\em Phys. Rev.
  Lett.}, vol.~121, p.~266601, Dec 2018.

\bibitem{Nonlinear-Interface2019}
D.~Choe, M.-J. Jin, S.-I. Kim, H.-J. Choi, J.~Jo, I.~Oh, J.~Park, H.~Jin, H.~C.
  Koo, B.-C. Min, S.~Hong, H.-W. Lee, S.-H. Baek, and J.-W. Yoo, ``Gate-tunable
  giant nonreciprocal charge transport in non-centrosymmetric oxide
  interfaces,'' {\em Nature Communications}, vol.~10, no.~1, p.~4510, 2019.

\bibitem{Dagan:2024}
G.~Tuvia, A.~Burshtein, I.~Silber, A.~Aharony, O.~Entin-Wohlman, M.~Goldstein,
  and Y.~Dagan, ``Enhanced nonlinear response by manipulating the Dirac point
  at the (111) ${\mathrm{LaTiO}}_{3}/{\mathrm{SrTiO}}_{3}$ interface,'' {\em
  Phys. Rev. Lett.}, vol.~132, p.~146301, Apr 2024.

\bibitem{Sodemann:2015}
I.~Sodemann and L.~Fu, ``Quantum nonlinear {H}all effect induced by {B}erry
  curvature dipole in time-reversal invariant materials,'' {\em Phys. Rev.
  Lett.}, vol.~115, p.~216806, Nov 2015.

\bibitem{Dzero:2024}
M.~Dzero, J.~Hasan, and A.~Levchenko, ``Resonant second harmonic generation in
  a two-dimensional electron system,'' arXiv:2411.10852 (2024).

\bibitem{Andrey2006}
A.~V. Shytov, E.~G. Mishchenko, H.-A. Engel, and B.~I. Halperin, ``Small-angle
  impurity scattering and the spin {H}all conductivity in two-dimensional
  semiconductor systems,'' {\em Phys. Rev. B}, vol.~73, p.~075316, Feb 2006.

\bibitem{Deyo:2009}
E.~Deyo, L.~E. Golub, E.~L. Ivchenko, and B.~Spivak, ``Semiclassical theory of
  the photogalvanic effect in non-centrosymmetric systems,'' 2009.

\bibitem{Edelshtein:1988}
V.~M. Edel'shtein, ``Second-harmonic generation in two-dimensional systems
  without inversion centers,'' {\em Sov. Phys. - JETP}, vol.~68, p.~1446, 1988.

\end{thebibliography}

\end{document}